\documentclass[english,12pt,a4paper]{article}
\usepackage{authblk}
\usepackage[text={17.0cm,23.7cm}]{geometry}
\usepackage[T1]{fontenc}
\usepackage[utf8]{inputenc}
\usepackage[english]{babel} 
\usepackage{hyperref}
\usepackage{graphicx}
\usepackage{amsmath}
\usepackage{amssymb}
\usepackage{siunitx} 
\usepackage[dvipsnames]{xcolor}
\usepackage{yfonts}
\usepackage{ulem}
\usepackage{enumitem}
\usepackage[mathscr]{euscript}
\usepackage[style=numeric-comp, sorting=none]{biblatex}
\usepackage{float}
\usepackage{amsmath}
\usepackage{csquotes}
\usepackage{color}
\usepackage{siunitx, physics} 

\newcommand*{\parsec}{\text{pc}}

\begin{document}
	
	\title{\vspace{-2cm}
		\vspace{0.6cm}
	\textbf{Dark matter explanations for the neutrino emission from the Seyfert galaxy NGC 1068}\\[8mm]}
\author[1]{Kensuke Akita}
\author[2]{Alejandro Ibarra}
\author[2,3]{Robert Zimmermann}

\affil[1]{\normalsize\textit{Department of Physics, The University of Tokyo, Bunkyo-ku, Tokyo 113-0033, Japan}}
\affil[2]{\normalsize\textit{Physik-Department, Technische Universit\"at M\"unchen, James-Franck-Stra\ss{}e, 85748 Garching, Germany}}
\affil[3]{\normalsize\textit{Max Planck Institute for Astrophysics, Karl-Schwarzschild-Straße 1, 85748 Garching, Germany}}

\date{}
\maketitle

\begin{abstract}
We investigate the possibility that the high-energy neutrino flux observed from the Seyfert galaxy NGC 1068 originates from dark matter annihilations within the density spike surrounding the supermassive black hole at its center. The comparatively lower gamma-ray flux is attributed to a dark sector that couples predominantly to Standard Model neutrinos. To explain the absence of a corresponding neutrino signal from the center of the Milky Way, we propose two scenarios: (i) the disruption of the dark matter spike at the Milky Way center due to stellar heating, or (ii) the annihilation into a dark scalar that decays exclusively into neutrinos, with a decay length longer than the size of the Milky Way but shorter than the distance from Earth to NGC 1068.
\end{abstract}

\section{Introduction}
\label{sec:Intro}

The IceCube Collaboration has reported a 4.2$\sigma$ excess of neutrinos with energies in the range of 1.5–15 TeV, spatially associated with the Seyfert galaxy NGC 1068~\cite{IceCube:2022der}. NGC 1068 is a well-studied Type II Seyfert galaxy that hosts an active galactic nucleus (AGN). In standard hadroproduction models of neutrino emission, one generally expects a photon flux comparable to the neutrino flux~\cite{Berezinsky:1975zz}. However, photon emission can be partially attenuated by pair production with ambient photons in the dense dust torus surrounding the central supermassive black hole. Despite this attenuation, the predicted high-energy gamma-ray flux exceeds the upper limits set by MAGIC~\cite{MAGIC:2019fvw} and Fermi-LAT~\cite{Fermi-LAT:2019yla,Ballet:2020hze}, unless the neutrinos originate from deep within the AGN, at distances of only 30–100 Schwarzschild radii from the black hole~\cite{Murase:2022dog}.

In this paper, we explore the possibility that the neutrino emission from NGC 1068, unaccompanied by a corresponding photon signal, originates from the annihilation of dark matter particles in its central region. This interpretation faces an immediate challenge: no similar neutrino excess has been observed from the Galactic Center of the Milky Way~\cite{IceCube:2023ies,KM3NeT:2024xca,Albert:2016emp}. Given that the Milky Way is only 8.2 kpc from Earth \cite{Balaji:2023hmy}, while NGC 1068 lies at a distance of 14.4 Mpc~\cite{Huang:2024yua}, explaining the IceCube detection would require a neutrino luminosity from NGC 1068 more than six orders of magnitude greater than that of the Milky Way. This seems puzzling, considering that both galaxies have comparable total masses.

However, both NGC 1068 and the Milky Way host a supermassive black hole (SMBH) at their centers. It has been shown that the adiabatic growth of such a black hole within a dark matter halo can induce the formation of a density ``spike'' where the dark matter annihilation rate is significantly enhanced~\cite{Quinlan:1994ed,Gondolo:1999ef, Sadeghian:2013laa,Ferrer:2017xwm,}. In the case of the Milky Way, this spike is believed to have been largely smoothed out due to gravitational heating from stars observed orbiting close to the central black hole~\cite{
Bertone:2005hw,Bertone:2005xv,Balaji:2023hmy}. In contrast, current instruments have not resolved such close-in stellar orbits around the black hole in NGC 1068.

In the following, we argue that if the dark matter spike in NGC 1068 is not severely affected by stellar heating, the resulting neutrino annihilation rate could be many orders of magnitude higher than in the Milky Way center, sufficient to account for the observed neutrino excess from NGC 1068 while remaining consistent with the absence of a signal from our own Galactic Center.

This work is organized as follows. In Section \ref{sec:spike} we describe the dark matter distribution in the center of NGC 1068 and in the Milky Way. In Section \ref{sec:neutrinos} we calculate the neutrino flux from these two targets assuming annihilations into a neutrino-antineutrino pair, and in Section \ref{sec:mediator} assuming annihilations into two scalar mediators that decay in flight into two neutrinos. Finally, in Section \ref{sec:conclusions} we present our conclusions.

\section{Density distribution in dark matter spikes}
\label{sec:spike}

The growth of primordial density perturbations in the early Universe and their subsequent collapse lead to dark matter halos with a density distribution that we assume to be described by the Navarro-Frenk-White (NFW) profile \cite{Navarro:1995iw,Navarro:1996gj}:
\begin{equation}
    \rho_{\rm NFW} (r) = \rho_s \left( \frac{r}{r_s} \right)^{-1} \left( 1 + \frac{r}{r_s} \right)^{-2} ~,
    \label{equ:nfw_fct}
\end{equation}
where $r_s$ is the scale radius and $\rho_s$ the scale density. This distribution is modified by the adiabatic growth of the supermassive black hole at the center of the galaxy, as well as by the possible annihilation of dark matter particles, resulting in a ``spike'' in the inner part of the distribution. The modified density profile, under the assumption that it is unperturbed, has the form~\cite{Quinlan:1994ed,Gondolo:1999ef}:
\begin{align}
    \rho_\psi (r) = 
    \begin{cases}
        \displaystyle{0} &\displaystyle{r < 4 R_{\rm S}}
        \\
        \displaystyle{\frac{\rho_{\rm sp}^{7/3}(r) \rho_{\rm c}}{\rho_{\rm sp}^{7/3}(r) + \rho_{\rm c}}}&\displaystyle{4 R_{\rm S} \leq r \leq R_{\rm sp}}
        \\
        \frac{\displaystyle{\rho_{\rm NFW}(r)} \rho_c}{\displaystyle{\rho_{\rm NFW}(r)} + \rho_c}&\displaystyle{r \geq R_{\rm sp}}
    \end{cases} ~,
    \label{equ:dm_energy_density_ngc}
\end{align}
with $R_{\rm S}=2 G M_{\rm BH}$ the Schwarzschild radius of the black hole, $R_{\rm sp}$ the size of the spike, and $\rho_{\rm sp}^\gamma(r)$ the density distribution in the absence of dark matter annihilations, given by:
\begin{equation}
    \rho_{\rm sp}^\gamma(r) = \rho_{\rm R} \left( 1 - \frac{4 R_{\rm S}}{r} \right)^3 \left( \frac{R_{\rm sp}}{r} \right)^\gamma ~,
    \label{equ:spike_fct}
\end{equation}
where $\rho_{\rm R}= \rho_s \left( R_{\rm sp}/r_s \right)^{-1} \left( 1 + R_{\rm sp}/r_s \right)^{-2} \left( 1 - 4 R_{\rm S}/R_{\rm sp} \right)^{-3} \simeq \rho_{s}\, (R_{\rm sp}/r_s)^{-1}$ is a normalization factor, chosen to match the density profile outside of the spike. Finally, $\rho_{\rm c}= m_\psi/(\langle \sigma v \rangle t_{\rm BH})$ is the maximum density of the spike in the presence of dark matter annihilations with $m_\psi$ the dark matter mass, $\langle \sigma v \rangle$ the thermally averaged annihilation cross-section, and $t_{\rm BH}$ the age of the SMBH.

The Milky Way also contains in its center a supermassive black hole which could have generated a dark matter spike. On the other hand, the stars orbiting very close to the black hole heat up the dark matter through gravitational interactions, thereby softening the spike~\cite{Bertone:2005hw,Bertone:2005xv}. When stellar heating is significant, the dark matter distribution in the central parts of the galaxy reads instead~\cite{Balaji:2023hmy}:
\begin{equation}
    \rho_\psi^{\rm MW} (r) = 
    \begin{cases}
        \displaystyle{0} &\displaystyle{r < 4 R_{\rm S}^{\rm MW}}
        \\
        \displaystyle{\frac{\rho_{\rm sp}^{3/2}(r) \rho_{\rm c}}{\rho_{\rm sp}^{3/2}(r) + \rho_{\rm c}}}&\displaystyle{4 R_{\rm S}^{\rm MW} \leq r \leq R_{\rm sp}^{\rm MW}}
        \\
        \frac{\displaystyle{\rho_{\rm NFW}(r)} \rho_c}{\displaystyle{\rho_{\rm NFW}(r)} + \rho_c}&\displaystyle{r \geq R_{\rm sp}^{\rm MW}}
    \end{cases} ~.
    \label{equ:dm_distribution_mw}
\end{equation}
In this paper, we will assume that NGC 1068 contains no stars close to its central supermassive black hole, so that its dark matter profile is described by Eq.~(\ref{equ:dm_energy_density_ngc}), while the dark matter profile of the Milky Way is described by Eq.~(\ref{equ:dm_distribution_mw}), due to the known existence of S-stars. For NGC 1068 we adopt $R_{\rm sp} = 0.7~\si{\kilo\parsec}$, $M_{\rm BH} = 10^7 \, M_\odot$, $r_h = 0.65~\si{\parsec}$, $\rho_s = 0.35~\si{\giga\electronvolt\per\centi\meter\cubed}$, $\gamma_{\rm sp}=7/3$, $r_s = 13~\si{\kilo\parsec}$, and $t_{\rm BH} = 10^9$ years \cite{Cline:2023tkp}, whereas for the Milky Way we take $\rho_s^{\rm MW} = \rho_\odot (R_\odot /r_s^{\rm MW}) (1+R_\odot / r_s^{\rm MW})^2 = 0.351~\si{\giga\electronvolt\per\centi\meter\cubed}$, $\rho_\odot = 0.383~\si{\giga\electronvolt\per\centi\meter\cubed}$, $R_\odot = 8.2~\si{\kilo\parsec}$, $r_s^{\rm MW}=18.6~\si{\kilo\parsec}$, $M_{\rm BH}^{\rm MW}= 4.3\times 10^6 \, M_\odot$, $R_{\rm sp}^{\rm MW}=0.34~\si{\parsec}$, $r_h^{\rm MW}=1.7~\si{\parsec}$, $\gamma_{\rm sp}^{\rm MW}=3/2$, and $t_{\rm BH}^{\rm MW}=10^{10}$ years \cite{Balaji:2023hmy}.
\par
\begin{figure}[t!]
    \centering
    \includegraphics[width=0.6\textwidth]{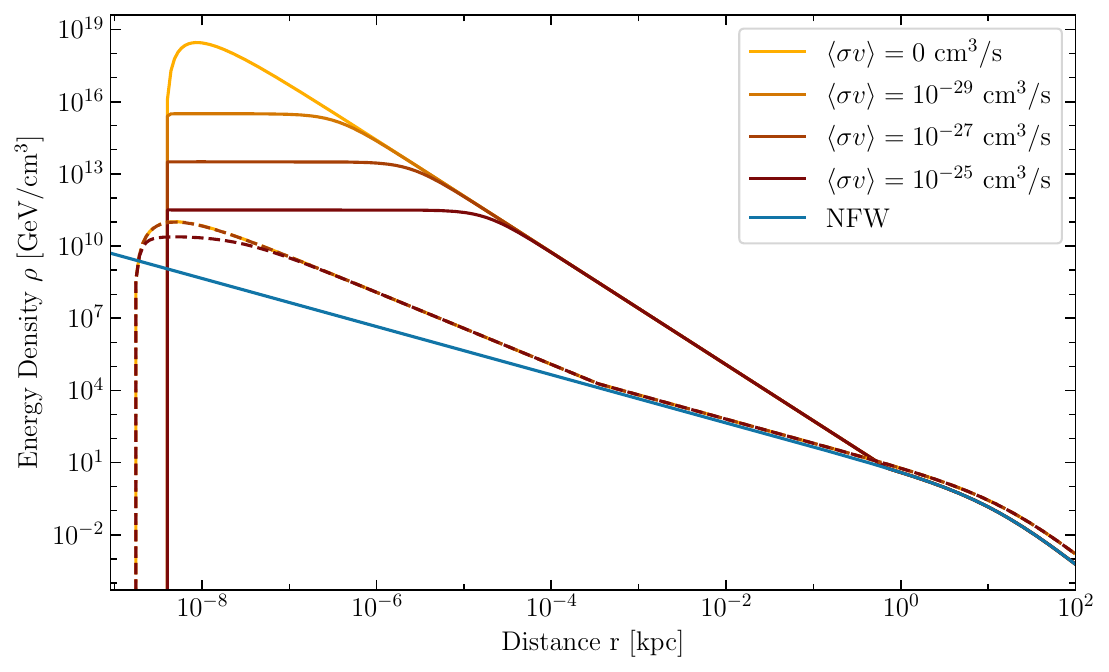}
    \caption{\small Dark Matter density distributions in the Milky Way (with stellar heating, dashed) and NGC 1068 (without stellar heating, solid) for different dark matter annihilation cross-sections and a dark matter mass $m_\psi = 1$ TeV. The NFW profile for NGC 1068 (blue solid) is shown for reference. }
    \label{fig:DM_spike_profile}
\end{figure}

In Fig. \ref{fig:DM_spike_profile} we show the DM distribution in the central parts of the Milky Way (dashed lines) and NGC 1068 (solid lines) for different values of the annihilation cross-section, assuming that the effects of stellar heating in this galaxy are negligible for NGC 1068. The dark matter mass is assumed to be $m_\psi = 1$ TeV. Further, we include the NFW profile of NGC 1068 (blue line) for reference. The spike of NGC 1068 is significantly larger and starts at a greater distance due to the bigger SMBH mass and the influence of stellar heating. We also see the flattening of the spike because of annihilations.

\section{Neutrino flux from NGC 1068 and the Milky Way center}
\label{sec:neutrinos}

We conservatively assume that the bulk of the neutrino emission in the direction of NGC 1068 originates from annihilations in the dark matter spike. Due to the small angular size of the spike compared to the resolution of IceCube, the differential flux of neutrinos plus antineutrinos can be simply calculated using the point source approximation:  
\begin{equation}
    \frac{d \phi^{\rm point}_{\nu + \bar{\nu}}}{dE_\nu} = \frac{1}{4 \pi d_L^2} \frac{\langle \sigma v \rangle}{2m_\psi^2} \frac{1}{3} \frac{dN_\nu}{dE_\nu}  \int_0^{R_{\rm sp}} dr \, 4 \pi r^2 \rho_{\rm DM}^2 (r) ~ ~,
    \label{equ:flux_direct_annihilation}
\end{equation}
where $\rho_{\rm DM}(r)$ is given in Eq.~(\ref{equ:dm_energy_density_ngc}) and $d_L$ is the distance from the source to the detector.~\footnote{We assume that dark matter is a self-conjugate particle, as common in the literature of indirect detection. If the dark matter particle and antiparticle can be distinguished, Eq.~(\ref{equ:flux_direct_annihilation}) should be divided by an additional factor of 2.}  

In the Milky Way center, the dark matter spike is largely softened by stellar heating. Therefore, the neutrino emission in the direction of the Galactic Center may not be dominated by the emission from annihilations in the spike, but instead by annihilations in the dark matter halo. The flux from annihilations in the halo in the solid angle $\Delta \Omega$, parameterized through the opening angle $\theta$ away from the Galactic Center, can be calculated with the line-of-sight integral:
\begin{equation}
    \frac{d \phi^{\rm l.o.s.}_{\nu + \bar{\nu}}}{dE_\nu} = \frac{1}{4 \pi} \frac{\langle \sigma v \rangle}{2m_\psi^2} \frac{1}{3} \frac{dN_\nu}{dE_\nu} \int_{\Delta\Omega} d\Omega \, \int_0^{R_{\rm max}}  \,  ds \, \rho_{\rm DM}^2(r) ~ ~,
    \label{equ:flux_direct_annihilation}
\end{equation}
where $\rho_{\rm DM}(r)$ is given in Eq.(\ref{equ:dm_distribution_mw}) and
\begin{equation}
r^2(s, \theta)=s^2+R_\odot^2-2 s \, R_\odot \cos \theta ~.
\end{equation}
In Eq.~(\ref{equ:flux_direct_annihilation}) we have truncated the line-of-sight integral to a finite value $R_{\rm max}$, to account for the finite size of the dark matter halo. In our numerical analysis, we will adopt $R_{\rm max}=\sqrt{R_{vir}^2-R_\odot^2 \sin^2\theta}+R_\odot \cos\theta$ with $R_{vir}=200~\si{\kilo\parsec}$ \cite{Balaji:2023hmy}, although our conclusions do not differ significantly on this choice. We find that for an opening angle $\theta=1^\circ$ ($0.1^\circ$) around the Galactic Center, the l.o.s.~contribution is a factor 82 (9) larger than using the point-source approximation of only the spike, implying that the spike gives only a subdominant contribution. This makes the detection of the spike at the Milky Way center as a point source extremely challenging, and instead the signal from annihilation from this target will be diffuse.

\begin{figure}[t!]
    \centering
    \includegraphics[width=0.49\textwidth]{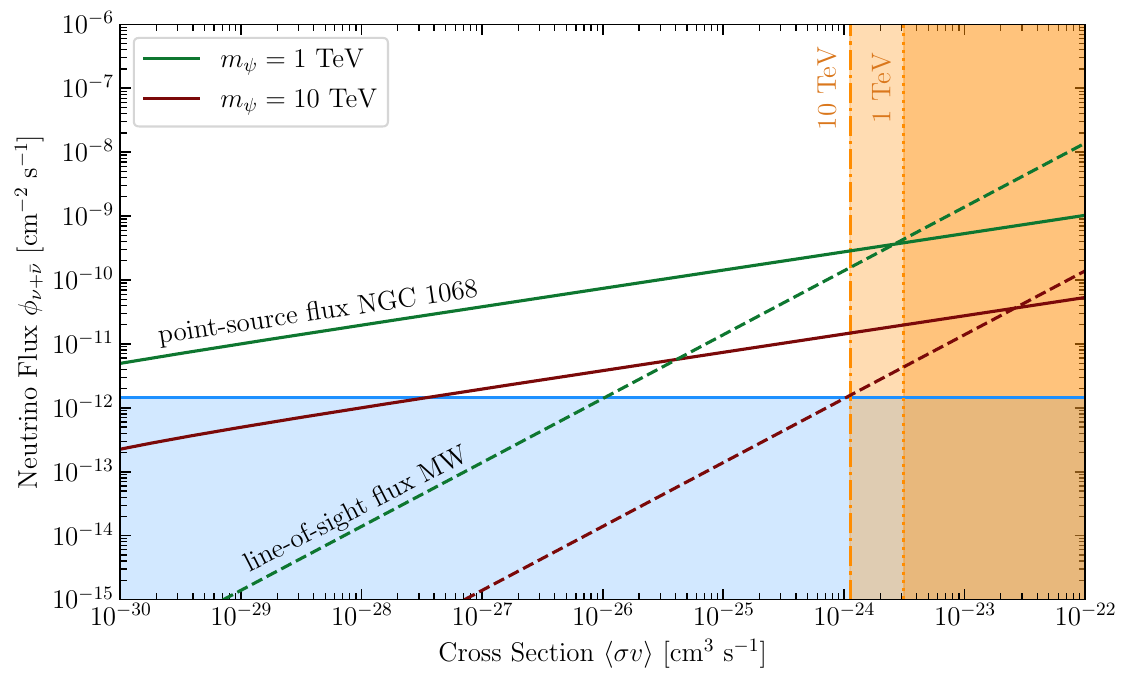}
    \includegraphics[width=0.49\textwidth]{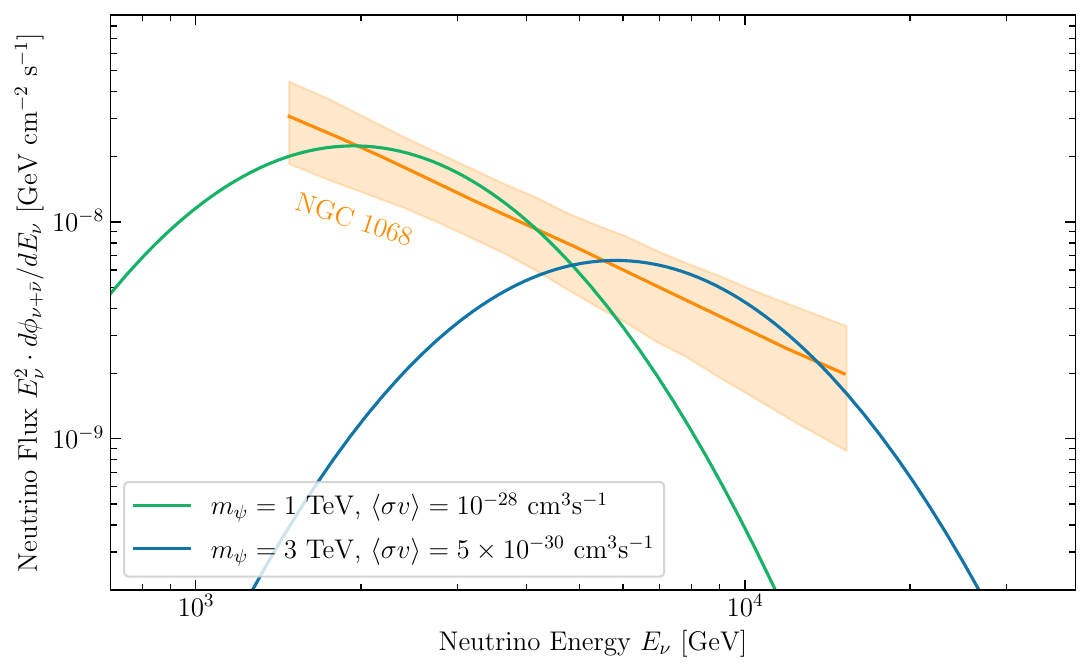}
    \caption{\small{ \it Left panel}: neutrino flux from the annihilation $\psi\psi\rightarrow \nu\bar\nu$ as a function of the annihilation cross-section from NGC 1068 (solid lines) and from a region of $\theta \leq1^\circ$ around the Galactic Center (dashed lines), assuming $m_\psi$=1 TeV (green line) or $m_\psi=10$ TeV (red lines). The blue shaded region correspond to the point-source fluxes that escape detection in IceCube in the direction of NGC 1068 \cite{IceCube:2021xar}, while the orange regions represent cross-sections excluded by the IceCube Collaboration from the non-observation of an exotic neutrino flux in the direction of the Milky Way center \cite{IceCube:2023ies}. {\it Right panel}: Neutrino energy spectrum of NGC 1068 at IceCube for two representative choices of the dark matter mass and annihilation cross-section. The orange band corresponds to the neutrino energy spectrum from NGC 1068 reconstructed by the IceCube Collaboration \cite{IceCube:2022der}.}
    \label{fig:point_source_direct}
\end{figure}

Let us first consider a scenario where dark matter annihilates into  $\nu\bar\nu$.
We show in Fig.\ref{fig:point_source_direct} the total neutrino flux from the spike of NGC 1068 (solid lines) and the diffuse flux in the direction of the Milky Way center assuming an opening angle of $1^\circ$ (dashed lines), as a function of the annihilation cross-section for a scenario where DM annihilates into $\nu \bar{\nu}$ for the specific choices $m_\psi=1$ TeV (green line) and 10 TeV (red lines). The orange band indicates the upper limit on the annihilation cross-section into $\nu\bar\nu$ from the non-observation of a signal from the Galactic Center at IceCube for $m_\psi=1$ and 10 TeV~\cite{IceCube:2023ies}. We also show in the plot as light blue the point-source sensitivity of IceCube in the direction of NGC 1068 (blue line) ~\cite{IceCube:2021xar}. When the annihilation cross-section is large, the flux from the Milky Way center is larger than from NGC 1068, and viceversa. However, for the values of the cross-section currently allowed by IceCube, the flux in the direction of NGC 1068 is expected to be larger. Furthermore, the expected flux is within the reach of the current point-source sensitivity of IceCube for $ \langle \sigma v\rangle\gtrsim 10^{-27}~{\rm cm^3}\,{\rm s}^{-1}$ if $m_\psi=10$ TeV and when $\langle \sigma v\rangle\gtrsim 3\times 10^{-31}~{\rm cm^3}\,{\rm s}^{-1}$ if $m_\psi=1$ TeV. This shows that if the spike in NGC 1068 is not affected by the stellar heating, dark matter annihilations could explain the neutrino emission from NGC 1068, while being compatible with the absence of a neutrino flux from the Milky Way center. It is also noteworthy that these values for the cross-section are in the ballpark of the values required by reproducing the observed dark matter abundance via thermal freeze-out.

Finally, we estimate the expected energy spectrum 
adopting an energy resolution of $ \sigma=0.25$ in $ \log_{10}E({\rm GeV}) $ around the peak position~\cite{Palomares-Ruiz:2007egs}. Therefore, the measured neutrino flux reads:
\begin{equation}
 \frac{d\phi_{\nu+\bar\nu}}{dE}\Big|_{\rm observed}=\int_0^{\infty}{\cal G}(E,E')\,\frac{d\phi_{\nu+\bar\nu}}{dE'}\,dE'\,, 
\end{equation}
where
\begin{equation}
 {\cal G}(E,\,E')=\frac{e^{-\frac{1}{2}(\sigma\ln10)^2}}{\sqrt{2\pi}\sigma\ln10}\,\frac{1}{E'}\,e^{-\frac{1}{2}\left( \frac{\log_{10}E/E'}{\sigma}\right) ^2}. 
\end{equation}
Two representative energy spectra are shown in Fig.~\ref{fig:point_source_direct}, compared to the energy spectrum of the neutrino emission from NGC 1068 reported by the IceCube Collaboration~\cite{IceCube:2022der}. Concretely, the green line corresponds to the annihilation $\psi\psi\rightarrow \nu\bar\nu$ with $m_\psi=1$ TeV and $\langle \sigma v\rangle= 10^{-28}\,{\rm cm}^3\,{\rm s}^{-1}$ and the blue line to $m_\psi=3$ TeV and $\langle \sigma v\rangle= 5\times 10^{-30}\,{\rm cm}^3\,{\rm s}^{-1}$. Both energy spectra are in rough agreement with the one reported by IceCube, although a dedicated simulation of our signal in IceCube would be necessary to determine the compatibility of the model with the data since the signal reconstruction is performed using a power law \cite{IceCube:2022der}. It is notable that IceCube has a high sensitivity to dark matter annihilations in NGC 1068, which is due to the large $J$-factor provoked by the spike. 

\section{Dark matter annihilation into a light mediator}
\label{sec:mediator}

We consider now annihilations into a dark sector scalar (or pseudoscalar) mediator $\phi$, that decays in flight $\phi\rightarrow\nu\bar{\nu}$
through a Yukawa coupling $g \phi\bar{\nu}\nu$~\cite{Berryman:2022hds}.\footnote{Similar conclusions hold for a vector mediator $V_\mu$ that couples to neutrinos through $g V_\mu \nu^\dagger\bar \sigma^\mu \nu$.}
If the mediator is sufficiently long-lived (in the galactic frame), it can decay outside of the Milky Way, thus leaving no neutrino signal from the Milky Way center. However, if the decay length is smaller than the distance from NGC 1068 to the Earth, a neutrino signal would be observed from this source. We will assume that the scalar  is light, so that the decays into charge particles is kinematically suppressed. Further, the small phase space available for the decay, combined with the large Lorentz factor imprinted on the sterile neutrino from the annihilation of heavy dark matter particles, translates into typically large decay lengths. 
More precisely, the decay length in the galactic frame reads:
\begin{align}
c\tau&\simeq \frac{8 \pi}{g^2}\frac{m_\psi}{m_\phi^2} \simeq 16\,{\rm kpc} ~ \Big(\frac{g}{10^{-12}} \Big)^{-2} \left( \frac{m_\psi}{1 ~{\rm TeV}} \right) \left( \frac{m_\phi}{100 ~{\rm keV}} \right)^{-2} .
\end{align}

\begin{figure}[t!]
    \centering
    \includegraphics[width=0.6\textwidth]{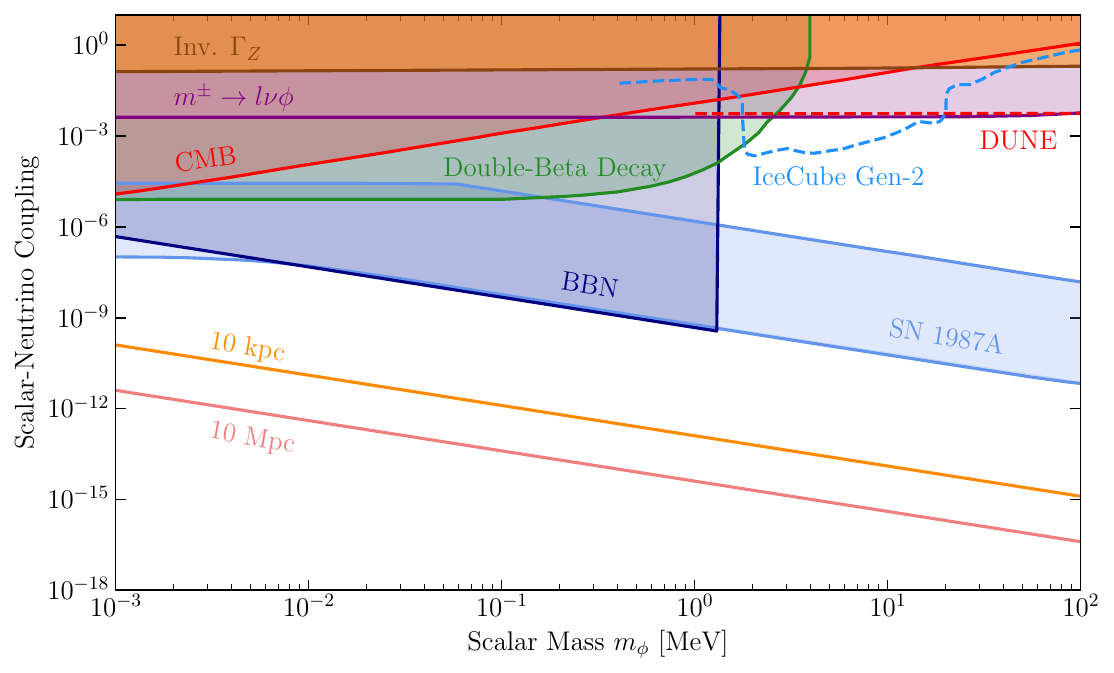}
    \caption{\small Values of the Yukawa coupling $g$ and mediator mass $m_\phi$ leading to a decay length of $\phi$ of 10 kpc and 10 Mpc from the dark matter annihilation $\psi\psi\rightarrow \phi\phi$ when $m_\psi=1$ TeV, The figure also shows various cosmological and laboratory constraints on the model \cite{Escudero:2019gvw,Farzan:2002wx,Heurtier:2016otg,Fiorillo:2022cdq,Akita:2023iwq,E949:2016gnh,deGouvea:2019qaz,Brune:2018sab}, as well as the projected sensitivities of IceCube Gen-2 \cite{IceCube-Gen2:2020qha} and DUNE \cite{Kelly:2019wow, Berryman:2018ogk}.}
    \label{fig:decay_length}
\end{figure}

We show in Fig.~\ref{fig:decay_length} the contours of the decay length of the scalar $\phi$ in the galactic frame, for $m_\psi=1$ TeV, in the parameter space spanned by the Yukawa coupling $g$ and the scalar mass $m_\phi$. The red line indicates a decay length of 10 Mpc, and the orange line 10 kpc. In the region in between, the neutrino flux from annihilations in the Galactic Center is suppressed, but not the one originating from NGC 1068. For comparison, we also show as a shaded area the region which is currently ruled out by Big Bang nucleosynthesis~\cite{Escudero:2019gvw}, SN~1987A \cite{Farzan:2002wx,Heurtier:2016otg,Fiorillo:2022cdq,Akita:2023iwq}, rare charged kaon decay~\cite{E949:2016gnh}, invisible Z boson decay~\cite{deGouvea:2019qaz}, and neutrinoless double beta decay with majoron emission~\cite{Brune:2018sab}, as well as the projected sensitivities of IceCube Gen-2 \cite{IceCube-Gen2:2020qha} and DUNE \cite{Kelly:2019wow, Berryman:2018ogk}. (for a compilation of constraints and a comprehensive review of neutrino ``secret interactions'', see \cite{Berryman:2022hds}).

\begin{figure}[t!]
    \centering
    \includegraphics[width=0.49\textwidth]{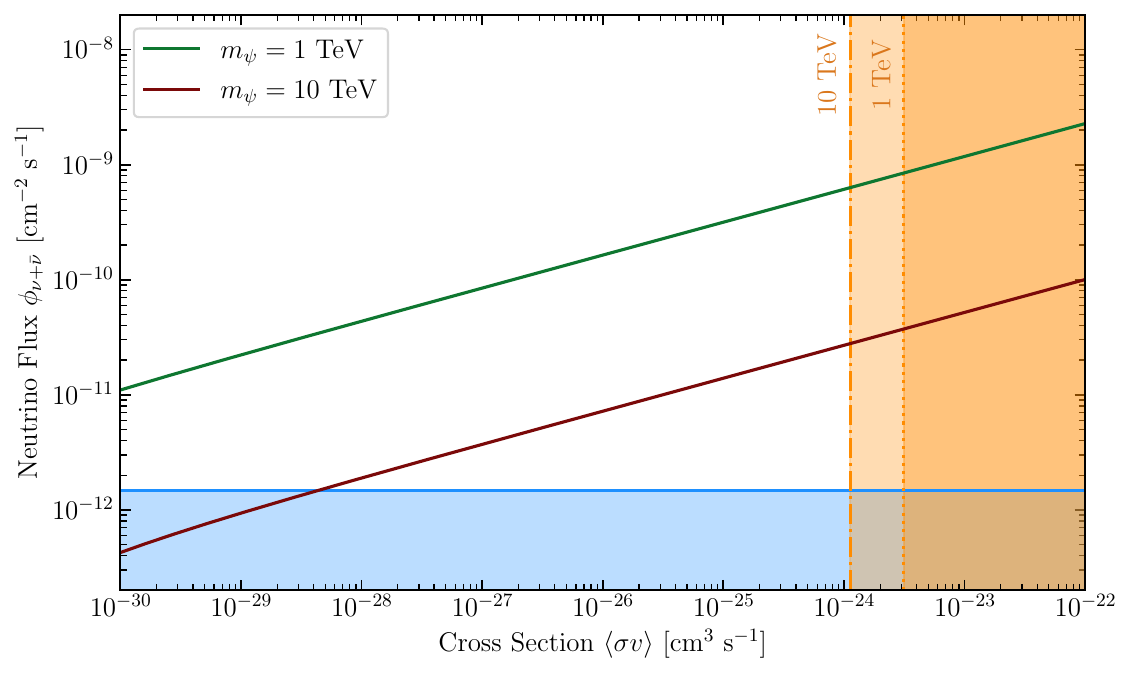}
    \includegraphics[width=0.49\textwidth]{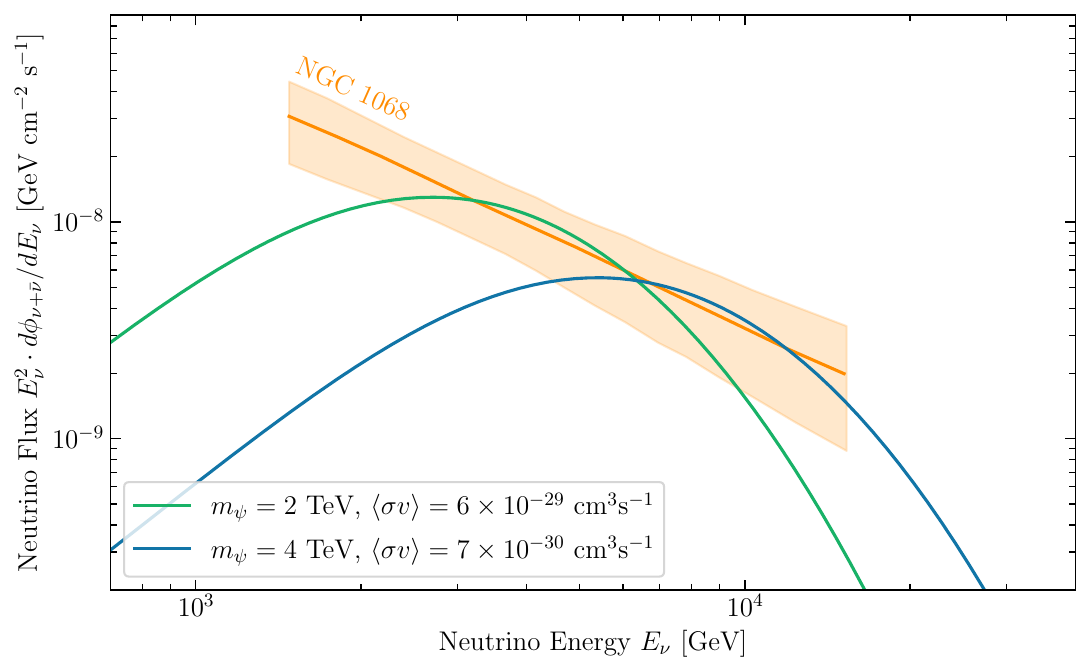}
    \caption{\small Same as Fig.~\ref{fig:point_source_direct}, but for the annihilation into a dark scalar $\psi\psi\rightarrow \phi\phi$, followed by the decay $\phi\rightarrow \nu\bar\nu$. For the plot we assumed $m_\phi=100$ keV and a Yukawa coupling $g=10^{-12}$, which results in a decay length larger than the distance between the Galactic Center and the Earth, but smaller than the distance between NGC 1068 and the Earth. }
    \label{fig:point_source_scalar}
\end{figure}

The energy spectrum of neutrinos decaying in flight resembles a box \cite{Ibarra:2012dw}, and is therefore less sharp than for direct annihilation $\psi\psi\rightarrow \nu\bar\nu$. Further, the higher multiplicity of neutrinos in the final state implies a larger flux at Earth. As an illustration, we show in Fig. \ref{fig:point_source_scalar} the total flux as a function of the cross-section for $g=10^{-12}$ and $m_\phi= 100$ keV. For these parameters, the decay length is $c\tau= 16~\si{\kilo\parsec}~ (160~\si{\kilo\parsec})$ for $m_\psi=1$ TeV (10 TeV), and the flux from the Milky Way is very suppressed as only a very small fraction of the scalar mediators decay in flight before reaching the detector.

Finally, let us briefly comment on an alternative scenario for long-lived mediators, consisting in the annihilation into sterile neutrinos, that then decay in flight into three neutrinos. The decay length is:
\begin{align}
c\tau&\simeq \frac{96 \pi^3}{G_{\rm F}^2} \frac{m_\psi}{m_N^6} \vert U_{lN}\vert^{-2}\simeq 74\,{\rm pc}\,
|U_{lN}|^{-2} \left( \frac{m_\psi}{1 ~{\rm TeV}} \right) \left( \frac{m_N}{1 ~{\rm MeV}} \right)^{-6} \;.
\end{align}
For $m_\psi=1$ TeV, the active-sterile mixing parameter and sterile neutrino mass such that the decay length is larger than 10 kpc and smaller than 10 Mpc are
\begin{align}
1.4 \times 10^{2} \lesssim 
|U_{lN}|^{-2}  \left( \frac{m_N}{1 ~{\rm MeV}} \right)^{-6} \lesssim 1.4 \times 10^{5}\;.
\end{align}
When assuming a minimal scenario with non-resonant production of right-handed neutrinos, this range of parameters is excluded by the combined measurements of the Hubble constant, Supernovae Ia luminosity distances, the Cosmic Microwave Background shift parameter, as well as measurements of the Baryon Acoustic Oscillation scale \cite{Vincent:2014rja}. Other scenarios with long-lived mediators could explain the large neutrino emission from NGC 1068, while being compatible with all existing laboratory and cosmological constraints.

\section{Conclusions}
\label{sec:conclusions}

We have proposed two possible explanations for the observed excess of neutrino emission from the Seyfert galaxy NGC 1068 relative to the expected gamma-ray flux. We argue that dark matter annihilations within a density spike surrounding the supermassive black hole at the center of NGC 1068 can produce a neutrino flux detectable by IceCube. The comparatively suppressed gamma-ray flux can be accounted for if the dark sector couples predominantly to Standard Model neutrinos.

A neutrino flux from the center of the Milky Way might also be expected, given its proximity to Earth. However, stellar heating by the S-stars orbiting close to the supermassive black hole at the Galactic Center can soften the dark matter density spike, significantly reducing the annihilation rate. In this work, we have shown that if the spike in NGC 1068 remains unaffected by such stellar heating, a neutrino point source could be observed in this direction for values of the annihilation cross-section that generate no observable neutrino flux in the direction of the Milky Way center.

An alternative explanation involves dark matter annihilating into two light dark scalars that subsequently decay in flight into a neutrino–antineutrino pair. If the scalar is sufficiently light, decays into charged particles are kinematically suppressed, thereby also suppressing the resulting photon flux. Moreover, if the coupling of the dark scalar to neutrinos is small, its decay length can be much larger than the distance from the Milky Way to the Earth, suppressing very strongly the neutrino flux from the Milky Way center. On the other hand, if the decay length is smaller than the distance from NGC 1068 to Earth, this galaxy could be detected in neutrinos.  Other extragalactic targets could similarly be explored, offering further insight into the intriguing possibility that IceCube may have detected neutrinos originating from dark matter annihilations.

\section*{Acknowledgments}
\noindent
We thank Elisa Resconi for useful discussions on the neutrino emission from NGC 1068, and Motoko Fujiwara for collaboration during the early stages of this work. 
This work is supported by the Collaborative Research Center SFB1258, by the Deutsche Forschungsgemeinschaft (DFG, German Research Foundation) under Germany's Excellence Strategy - EXC-2094 - 390783311, the JSPS Grant-in-Aid for Scientific Research KAKENHI Grant No.~24KJ0060, and the European Union-funded project \texttt{mw-atlas} under grant agreement No. 101166905.

\printbibliography

@article{Bertone:2005hw,
    author = "Bertone, Gianfranco and Merritt, David",
    title = "{Time-dependent models for dark matter at the Galactic Center}",
    eprint = "astro-ph/0501555",
    archivePrefix = "arXiv",
    reportNumber = "FERMILAB-PUB-05-013-A",
    doi = "10.1103/PhysRevD.72.103502",
    journal = "Phys. Rev. D",
    volume = "72",
    pages = "103502",
    year = "2005"
}

@article{deGouvea:2019qaz,
    author = "de Gouv{\^e}a, Andr{\'e} and Dev, P. S. Bhupal and Dutta, Bhaskar and Ghosh, Tathagata and Han, Tao and Zhang, Yongchao",
    title = "{Leptonic Scalars at the LHC}",
    eprint = "1910.01132",
    archivePrefix = "arXiv",
    primaryClass = "hep-ph",
    reportNumber = "PITT-PACC 1909, MI-TH-1936",
    doi = "10.1007/JHEP07(2020)142",
    journal = "JHEP",
    volume = "07",
    pages = "142",
    year = "2020"
}

@article{Vincent:2014rja,
    author = "Vincent, Aaron C. and Martinez, Enrique Fernandez and Hern{\'a}ndez, Pilar and Lattanzi, Massimiliano and Mena, Olga",
    title = "{Revisiting cosmological bounds on sterile neutrinos}",
    eprint = "1408.1956",
    archivePrefix = "arXiv",
    primaryClass = "astro-ph.CO",
    reportNumber = "IFIC-14-53, FTUAM-14-32, IFT-UAM-CSIC-14-075",
    doi = "10.1088/1475-7516/2015/04/006",
    journal = "JCAP",
    volume = "04",
    pages = "006",
    year = "2015"
}

@article{Ibarra:2012dw,
    author = "Ibarra, Alejandro and Lopez Gehler, Sergio and Pato, Miguel",
    title = "{Dark matter constraints from box-shaped gamma-ray features}",
    eprint = "1205.0007",
    archivePrefix = "arXiv",
    primaryClass = "hep-ph",
    doi = "10.1088/1475-7516/2012/07/043",
    journal = "JCAP",
    volume = "07",
    pages = "043",
    year = "2012"
}

@article{Palomares-Ruiz:2007egs,
    author = "Palomares-Ruiz, Sergio",
    title = "{Model-independent bound on the dark matter lifetime}",
    eprint = "0712.1937",
    archivePrefix = "arXiv",
    primaryClass = "astro-ph",
    reportNumber = "IPPP-07-96, DCPT-07-192",
    doi = "10.1016/j.physletb.2008.05.040",
    journal = "Phys. Lett. B",
    volume = "665",
    pages = "50--53",
    year = "2008"
}

@article{Bertone:2005xv,
    author = "Bertone, Gianfranco and Merritt, David",
    title = "{Dark matter dynamics and indirect detection}",
    eprint = "astro-ph/0504422",
    archivePrefix = "arXiv",
    reportNumber = "FERMILAB-PUB-05-051-A",
    doi = "10.1142/S0217732305017391",
    journal = "Mod. Phys. Lett. A",
    volume = "20",
    pages = "1021",
    year = "2005"
}

@article{Balaji:2023hmy,
    author = "Balaji, Shyam and Sachdeva, Divya and Sala, Filippo and Silk, Joseph",
    title = "{Dark matter spikes around Sgr A* in \ensuremath{\gamma}-rays}",
    eprint = "2303.12107",
    archivePrefix = "arXiv",
    primaryClass = "hep-ph",
    doi = "10.1088/1475-7516/2023/08/063",
    journal = "JCAP",
    volume = "08",
    pages = "063",
    year = "2023"
}

@article{Berryman:2022hds,
    author = "Berryman, Jeffrey M. and others",
    title = "{Neutrino self-interactions: A white paper}",
    eprint = "2203.01955",
    archivePrefix = "arXiv",
    primaryClass = "hep-ph",
    reportNumber = "CERN-TH-2022-024, DESY-22-035, FERMILAB-PUB-22-099-T",
    doi = "10.1016/j.dark.2023.101267",
    journal = "Phys. Dark Univ.",
    volume = "42",
    pages = "101267",
    year = "2023"
}

@article{IceCube:2023ies,
    author = "Abbasi, R. and others",
    collaboration = "IceCube",
    title = "{Search for neutrino lines from dark matter annihilation and decay with IceCube}",
    eprint = "2303.13663",
    archivePrefix = "arXiv",
    primaryClass = "astro-ph.HE",
    doi = "10.1103/PhysRevD.108.102004",
    journal = "Phys. Rev. D",
    volume = "108",
    number = "10",
    pages = "102004",
    year = "2023"
}

@article{KM3NeT:2024xca,
    author = "Aiello, S. and others",
    collaboration = "KM3NeT",
    title = "{First searches for dark matter with the KM3NeT neutrino telescopes}",
    eprint = "2411.10092",
    archivePrefix = "arXiv",
    primaryClass = "astro-ph.HE",
    doi = "10.1088/1475-7516/2025/03/058",
    journal = "JCAP",
    volume = "03",
    pages = "058",
    year = "2025"
}

@article{Sadeghian:2013laa,
    author = "Sadeghian, Laleh and Ferrer, Francesc and Will, Clifford M.",
    title = "{Dark matter distributions around massive black holes: A general relativistic analysis}",
    eprint = "1305.2619",
    archivePrefix = "arXiv",
    primaryClass = "astro-ph.GA",
    doi = "10.1103/PhysRevD.88.063522",
    journal = "Phys. Rev. D",
    volume = "88",
    number = "6",
    pages = "063522",
    year = "2013"
}

@article{Ferrer:2017xwm,
    author = "Ferrer, Francesc and da Rosa, Augusto Medeiros and Will, Clifford M.",
    title = "{Dark matter spikes in the vicinity of Kerr black holes}",
    eprint = "1707.06302",
    archivePrefix = "arXiv",
    primaryClass = "astro-ph.CO",
    doi = "10.1103/PhysRevD.96.083014",
    journal = "Phys. Rev. D",
    volume = "96",
    number = "8",
    pages = "083014",
    year = "2017"
}

@article{Albert:2016emp,
    author = "Albert, A. and others",
    title = "{Results from the search for dark matter in the Milky Way with 9 years of data of the ANTARES neutrino telescope}",
    eprint = "1612.04595",
    archivePrefix = "arXiv",
    primaryClass = "astro-ph.HE",
    doi = "10.1016/j.physletb.2017.03.063",
    journal = "Phys. Lett. B",
    volume = "769",
    pages = "249--254",
    year = "2017",
    note = "[Erratum: Phys.Lett.B 796, 253--255 (2019)]"
}

@article{IceCube:2022der,
    author = "Abbasi, R. and others",
    collaboration = "IceCube",
    title = "{Evidence for neutrino emission from the nearby active galaxy NGC 1068}",
    eprint = "2211.09972",
    archivePrefix = "arXiv",
    primaryClass = "astro-ph.HE",
    doi = "10.1126/science.abg3395",
    journal = "Science",
    volume = "378",
    number = "6619",
    pages = "538--543",
    year = "2022"
}

@article{Murase:2022dog,
    author = "Murase, Kohta",
    title = "{Hidden Hearts of Neutrino Active Galaxies}",
    eprint = "2211.04460",
    archivePrefix = "arXiv",
    primaryClass = "astro-ph.HE",
    doi = "10.3847/2041-8213/aca53c",
    journal = "Astrophys. J. Lett.",
    volume = "941",
    number = "1",
    pages = "L17",
    year = "2022"
}

@article{Ballet:2020hze,
    author = "Ballet, J. and Burnett, T. H. and Digel, S. W. and Lott, B.",
    collaboration = "Fermi-LAT",
    title = "{Fermi Large Area Telescope Fourth Source Catalog Data Release 2}",
    eprint = "2005.11208",
    archivePrefix = "arXiv",
    primaryClass = "astro-ph.HE",
    month = "5",
    year = "2020"
}

@article{Fermi-LAT:2019yla,
    author = "Abdollahi, S. and others",
    collaboration = "Fermi-LAT",
    title = "{$Fermi$ Large Area Telescope Fourth Source Catalog}",
    eprint = "1902.10045",
    archivePrefix = "arXiv",
    primaryClass = "astro-ph.HE",
    doi = "10.3847/1538-4365/ab6bcb",
    journal = "Astrophys. J. Suppl.",
    volume = "247",
    number = "1",
    pages = "33",
    year = "2020"
}

@article{MAGIC:2019fvw,
    author = "Acciari, V. A. and others",
    collaboration = "MAGIC",
    title = "{Constraints on gamma-ray and neutrino emission from NGC 1068 with the MAGIC telescopes}",
    eprint = "1906.10954",
    archivePrefix = "arXiv",
    primaryClass = "astro-ph.HE",
    doi = "10.3847/1538-4357/ab3a51",
    journal = "Astrophys. J.",
    volume = "883",
    pages = "135",
    year = "2019"
}

@article{Berezinsky:1975zz,
    author = "Berezinsky, V. S. and Smirnov, A. Yu.",
    title = "{Cosmic neutrinos of ultra-high energies and detection possibility}",
    doi = "10.1007/BF00643157",
    journal = "Astrophys. Space Sci.",
    volume = "32",
    pages = "461--482",
    year = "1975"
}

@article{Navarro:1996gj,
    author = "Navarro, Julio F. and Frenk, Carlos S. and White, Simon D. M.",
    title = "{A Universal density profile from hierarchical clustering}",
    eprint = "astro-ph/9611107",
    archivePrefix = "arXiv",
    doi = "10.1086/304888",
    journal = "Astrophys. J.",
    volume = "490",
    pages = "493--508",
    year = "1997"
}

@article{Navarro:1995iw,
    author = "Navarro, Julio F. and Frenk, Carlos S. and White, Simon D. M.",
    title = "{The Structure of cold dark matter halos}",
    eprint = "astro-ph/9508025",
    archivePrefix = "arXiv",
    doi = "10.1086/177173",
    journal = "Astrophys. J.",
    volume = "462",
    pages = "563--575",
    year = "1996"
}

@article{Quinlan:1994ed,
    author = "Quinlan, Gerald D. and Hernquist, Lars and Sigurdsson, Steinn",
    title = "{Models of Galaxies with Central Black Holes: Adiabatic Growth in Spherical Galaxies}",
    eprint = "astro-ph/9407005",
    archivePrefix = "arXiv",
    doi = "10.1086/175295",
    journal = "Astrophys. J.",
    volume = "440",
    pages = "554--564",
    year = "1995"
}

@article{Gondolo:1999ef,
    author = "Gondolo, Paolo and Silk, Joseph",
    title = "{Dark matter annihilation at the galactic center}",
    eprint = "astro-ph/9906391",
    archivePrefix = "arXiv",
    reportNumber = "MPI-PHT-99-10, OUAST-99-9",
    doi = "10.1103/PhysRevLett.83.1719",
    journal = "Phys. Rev. Lett.",
    volume = "83",
    pages = "1719--1722",
    year = "1999"
}

@article{Cline:2023tkp,
    author = "Cline, James M. and Puel, Matteo",
    title = "{NGC 1068 constraints on neutrino-dark matter scattering}",
    eprint = "2301.08756",
    archivePrefix = "arXiv",
    primaryClass = "hep-ph",
    doi = "10.1088/1475-7516/2023/06/004",
    journal = "JCAP",
    volume = "06",
    pages = "004",
    year = "2023"
}

@article{Huang:2024yua,
    author = "Huang, Yong-Han and Wang, Kai and Ma, Zhi-Peng",
    title = "{High-energy Neutrino Emission from NGC 1068 by Outflow-cloud Interactions}",
    eprint = "2406.14001",
    archivePrefix = "arXiv",
    primaryClass = "astro-ph.HE",
    month = "6",
    year = "2024"
}

@article{IceCube:2021xar,
    author = "Abbasi, R. and others",
    collaboration = "IceCube",
    title = "{IceCube Data for Neutrino Point-Source Searches Years 2008-2018}",
    eprint = "2101.09836",
    archivePrefix = "arXiv",
    primaryClass = "astro-ph.HE",
    doi = "10.21234/CPKQ-K003",
    month = "1",
    year = "2021"
}

@article{Escudero:2019gvw,
    author = "Escudero, Miguel and Witte, Samuel J.",
    title = "{A CMB search for the neutrino mass mechanism and its relation to the Hubble tension}",
    eprint = "1909.04044",
    archivePrefix = "arXiv",
    primaryClass = "astro-ph.CO",
    reportNumber = "KCL-2019-71",
    doi = "10.1140/epjc/s10052-020-7854-5",
    journal = "Eur. Phys. J. C",
    volume = "80",
    number = "4",
    pages = "294",
    year = "2020"
}

@article{E949:2016gnh,
    author = "Artamonov, A. V. and others",
    collaboration = "E949",
    title = "{Search for the rare decay $K^+\to\mu^+\nu\bar\nu\nu$}",
    eprint = "1606.09054",
    archivePrefix = "arXiv",
    primaryClass = "hep-ex",
    reportNumber = "FERMILAB-PUB-16-364-CD",
    doi = "10.1103/PhysRevD.94.032012",
    journal = "Phys. Rev. D",
    volume = "94",
    number = "3",
    pages = "032012",
    year = "2016"
}

@article{Brune:2018sab,
    author = {Brune, Tim and P{\"a}s, Heinrich},
    title = "{Massive Majorons and constraints on the Majoron-neutrino coupling}",
    eprint = "1808.08158",
    archivePrefix = "arXiv",
    primaryClass = "hep-ph",
    reportNumber = "DO-TH 18/23",
    doi = "10.1103/PhysRevD.99.096005",
    journal = "Phys. Rev. D",
    volume = "99",
    number = "9",
    pages = "096005",
    year = "2019"
}

@article{Akita:2023iwq,
    author = "Akita, Kensuke and Im, Sang Hui and Masud, Mehedi and Yun, Seokhoon",
    title = "{Limits on heavy neutral leptons, Z$^{\prime}$ bosons and majorons from high-energy supernova neutrinos}",
    eprint = "2312.13627",
    archivePrefix = "arXiv",
    primaryClass = "hep-ph",
    reportNumber = "CTPU-PTC-23-55",
    doi = "10.1007/JHEP07(2024)057",
    journal = "JHEP",
    volume = "07",
    pages = "057",
    year = "2024"
}

@article{Fiorillo:2022cdq,
    author = "Fiorillo, Damiano F. G. and Raffelt, Georg G. and Vitagliano, Edoardo",
    title = "{Strong Supernova 1987A Constraints on Bosons Decaying to Neutrinos}",
    eprint = "2209.11773",
    archivePrefix = "arXiv",
    primaryClass = "hep-ph",
    doi = "10.1103/PhysRevLett.131.021001",
    journal = "Phys. Rev. Lett.",
    volume = "131",
    number = "2",
    pages = "021001",
    year = "2023"
}

@article{Farzan:2002wx,
    author = "Farzan, Yasaman",
    title = "{Bounds on the coupling of the Majoron to light neutrinos from supernova cooling}",
    eprint = "hep-ph/0211375",
    archivePrefix = "arXiv",
    reportNumber = "SLAC-PUB-9543, SISSA-69-2002-EP",
    doi = "10.1103/PhysRevD.67.073015",
    journal = "Phys. Rev. D",
    volume = "67",
    pages = "073015",
    year = "2003"
}

@article{Heurtier:2016otg,
    author = "Heurtier, Lucien and Zhang, Yongchao",
    title = "{Supernova Constraints on Massive (Pseudo)Scalar Coupling to Neutrinos}",
    eprint = "1609.05882",
    archivePrefix = "arXiv",
    primaryClass = "hep-ph",
    reportNumber = "ULB-TH-16-16",
    doi = "10.1088/1475-7516/2017/02/042",
    journal = "JCAP",
    volume = "02",
    pages = "042",
    year = "2017"
}

@article{IceCube-Gen2:2020qha,
    author = "Aartsen, M. G. and others",
    collaboration = "IceCube-Gen2",
    title = "{IceCube-Gen2: the window to the extreme Universe}",
    eprint = "2008.04323",
    archivePrefix = "arXiv",
    primaryClass = "astro-ph.HE",
    doi = "10.1088/1361-6471/abbd48",
    journal = "J. Phys. G",
    volume = "48",
    number = "6",
    pages = "060501",
    year = "2021"
}

@article{Kelly:2019wow,
    author = "Kelly, Kevin J. and Zhang, Yue",
    title = "{Mononeutrino at DUNE: New Signals from Neutrinophilic Thermal Dark Matter}",
    eprint = "1901.01259",
    archivePrefix = "arXiv",
    primaryClass = "hep-ph",
    reportNumber = "FERMILAB-PUB-19-002-T",
    doi = "10.1103/PhysRevD.99.055034",
    journal = "Phys. Rev. D",
    volume = "99",
    number = "5",
    pages = "055034",
    year = "2019"
}

@article{Berryman:2018ogk,
    author = "Berryman, Jeffrey M. and De Gouv{\^e}a, Andr{\'e} and Kelly, Kevin J. and Zhang, Yue",
    title = "{Lepton-Number-Charged Scalars and Neutrino Beamstrahlung}",
    eprint = "1802.00009",
    archivePrefix = "arXiv",
    primaryClass = "hep-ph",
    reportNumber = "NUHEP-TH-18-03, FERMILAB-PUB-18-020-T",
    doi = "10.1103/PhysRevD.97.075030",
    journal = "Phys. Rev. D",
    volume = "97",
    number = "7",
    pages = "075030",
    year = "2018"
}

\end{document}